\documentclass[draft]{agujournal2019}
\journalname{Physical Review E}

\usepackage{graphicx} 
\usepackage{subfig}
\usepackage{dblfloatfix} 
\usepackage{soul}
\usepackage{lineno}

\begin{document}

\title{The role of injection method on residual trapping at the pore-scale in continuum-scale samples}


\authors{Catherine Spurin\affil{1}, Sharon Ellman\affil{2}, Tom Bultreys\affil{2}, and Hamdi Tchelepi\affil{1}}

\affiliation{1}{Department of Energy Science \& Engineering, Stanford University}
\affiliation{2}{Department of Geology, Ghent}




\begin{abstract}
The injection of CO$_2$ into underground reservoirs provides a long term solution for anthropogenic emissions. A variable injection method (such as ramping the flow rate up or down) provides flexibility to injection sites, and could increase trapping at the pore-scale. However, the impact of a variable injection method on the connectivity of the gas, and subsequent trapping has not been explored at the pore-scale. Here, we conduct pore-scale imaging in a continuum-scale sample to observe the role of a variable flow rate on residual trapping. We show that the injection method influences how much of the pore space is accessible to the gas, even when total volumes injected, and total flow rates remain constant. Starting at a high flow rate, then decreasing it, leads to a larger amount of the pore space accessed by the gas. Conversely, starting at a low flow rate, and increasing it, leads to a larger role of heterogeneity of the pore space. This can promote trapping efficiency because channelling of the two fluids can occur, but less gas is trapped overall. Overall, a high-to-low injection scenario is optimum for residual trapping in the pore space due to increases in pore space accessibility. 
\end{abstract}




\section{Introduction}
The subsurface storage of CO$_2$ will be integral to mitigating climate change, and limiting warming to the 2$^\circ$C agreed under the Paris agreement \cite{rubin2005ipcc, ipcc2014mitigation}. The trapping of CO$_2$ in the subsurface over geological time scales can be achieved in many ways: structural trapping, residual trapping, solubility trapping and mineral trapping \cite{krevor2015capillary, benson2012carbon}. Residual trapping is an important trapping mechanism as it can trap large volumes of CO$_2$ over short timescales \cite{krevor2015capillary}. It also limits the spread of the plume (which reduces the area that monitoring has to be conducted over), and is not dependent on the structural integrity of a cap rock. Storage projects require flexibility in variables such as CO$_2$ injection rate, with flow regularly ramped up or down. Variations in injection rate will influence the pore-scale movement of CO$_2$, and could play a significant role on the amount of residual trapping. There is also the potential to engineer injection strategies to maximise trapping \cite{shamshiri2012controlled}.

The role of the injection rate on trapping has been studied widely, over many scales. Pore-scale experiments have explored the role of flow rate on trapping, with a lower flow rate favouring snap-off of gas phase, which encourages trapping \cite{hughes2000pore, herring2015efficiently, wildenschild2001flow}. A higher flow rate causes a higher capillary pressure, allowing more of the pore space to be accessed by the gas \cite{bluntbook}. A higher gas saturation promotes a higher trapping efficiency \cite{land1968calculation, niu2015impact}.  \citeA{zhang2023impact} found that higher injection rates increased the migration distances and trapping capacities. Contrary to pore-scale studies, field scale models do not predict strong injection variability effects. For example, \citeA{bannach2015stable} found that varying CO$_2$ injection rates had a negative effect on overall injectivity within the first years of operation, but the role of the injection rate was deemed insignificant over longer periods. \citeA{li2019effects} found that varying the injection method improved residual trapping by up to 15\%, but only in the first years of injection. The difference in trapping efficiencies among the injection scenarios were within a few percent in the long term. \citeA{kolster2018impact} found that varying the amplitude and frequency of CO$_2$ storage had little long term impact (over 100 years) on reservoir pressure and plume migration. Ramping up the injection rate did not influence trapping, and so injection wells were advised to be placed as needed and progressively to avoid a large upfront cost of deployment to meet future storage demand. 

Overall, the literature is ambiguous on the impact of injection rate for CO$_2$ trapping, especially at the field scale. In these models, it appears that the potential impact is over short timescales, and is not important for long term trapping. However, the early stage dynamics will influence plume shape and later scale dynamics \cite{szulczewski2012lifetime, macminn2010co2}. Thus, a deeper understanding is required to suitably model plume injection and CO$_2$ trapping. If plume geometry and pressure distribution are poorly predicted it will affect estimates of storage capacity and residual trapping.

In this work we explore the effect of variable injection rates on pore-scale trapping in a continuum-scale sample. We compare the flow mechanisms generated when the flow rate is either ramped up, or ramped down, and explore the subsequent impact on trapping efficiency and utilisation of the pore space. With this analysis, we gain a deeper insight on the role of the injection scenario on residual trapping which will aid future modelling efforts, and bridge the gap between pore-scale observations, and their impact on larger scale flow properties.

\section{Material \& Methods}
\subsection{Experimental Apparatus and Procedure}
The experiments were conducted in a cylindrical Bentheimer sandstone sample, 25 mm in diameter and 45 mm in length. Details of the Hassler-type core holder used in these experiments is discussed elsewhere \cite{wang2023pore}. The sample was initially saturated with brine (deionized water doped with 17 wt.\% KI to improve the X-ray contrast). The system was pressurized to 5 MPa to minimise the compressibility of nitrogen, with an additional 2 MPa of confining pressure. For both experiments, nitrogen is initially injected (the drainage step), after which brine is injected (the imbibition step). There are two injection scenarios, during drainage, discussed in this work: high to low (referred herein as H2L) and low to high (referred herein as L2H). The flow rate during the high flow rate period was 1.8ml/min for 1hr. The flow rate during the low flow rate period was 0.6ml/min for 1hr. The flow rate during imbibition was constant at 1.2ml/min for 8 minutes. 

The total volume of gas injected was kept constant at 144ml (108ml during high flow, and 36ml during low flow). The volume of brine injected was kept constant at 10 ml. For this sample, 1 pore volume is approximately 5ml. This means around 20 pore volumes (PVs) were injected during high flow, 7 PVs were injected during low flow, and 2 PVs were injected during imbibition. 

\subsection{X-ray imaging}
The X-ray imaging was performed with the “High Energy Micro-CT Optimized for Research” scanner (HECTOR) at the center for X-ray tomography at Ghent University. The sample was exposed to X-ray radiation with a peak energy of 160 keV. Each tomogram contained 2101 projections. Two scans were required to capture the length of the core, with each scan lasting 20 minutes. One set of scans were taken during high flow, and also during low flow, at the end of drainage, and at the end of imbibition, resulting in 8 scans for the 4 stages of each injection scenario. 

\subsection{Image Analysis}
\begin{figure}[!htb]
	\begin{center}
		\centering
		\includegraphics*[angle=0, width=0.7\linewidth]{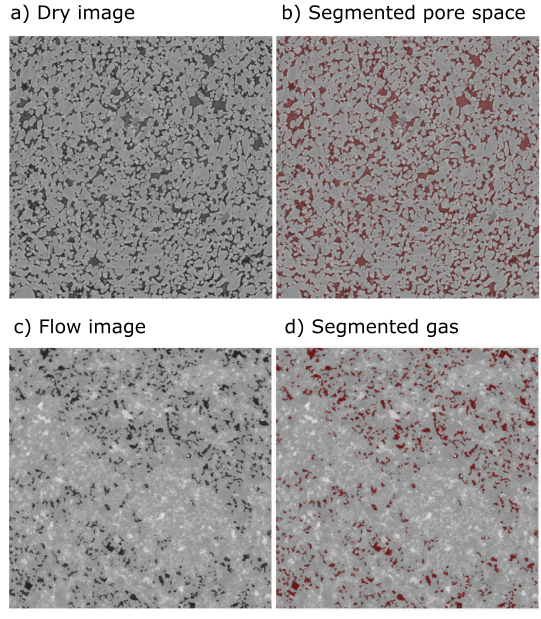}
	\end{center}
	\caption{a) the dry/ air-saturated image, b) the segmented pore space overlain on the dry/ air-saturated image, c) the flow image, with gas and water present in the pore space, and d) the segmented gas phase overlain on the flow image. }
	\label{processing_workflow}
\end{figure}

The voxel size for the images acquired was 20 $\mu m$. The images analyzed were $25 \times 25 \times 38$~mm in size (this means that dynamics do occur outside the field of view in the direction of flow, but the entire cross section of the core was imaged). The images were reconstructed from the X-ray projections, then filtered with a non-local means filter to suppress noise, while maintaining the information of phase boundaries \cite{schluter2014image}. Prior to the flow experiments, an image with only air in the pore space was taken. This image is used to segment the pore space from the rock grains using a watershed segmentation algorithm. All subsequent images are registered to this image. Then the sample was saturated with brine and imaged, with all subsequent images with the gas and brine present subtracted from this image to locate the gas. From this, the gas is segmented using a simple greyscale value threshold. The pore space was overlain on this segmentation to locate the pore space occupied with brine. The segmentation of the pore space is shown in Figure \ref{processing_workflow} a-b, and the segmentation of the gas is shown in Figure \ref{processing_workflow} c-d.

The image analysis workflow is available on Github: \url{https://github.com/cspurin/image_processing.git}, and described in detail in the Supplementary Information.

\section{Results and Discussion}
\subsection{The impact of injection scenario on saturation during drainage}
\begin{figure}[!htb]
	\begin{center}
		\centering
		\includegraphics*[angle=0, width=0.7\linewidth]{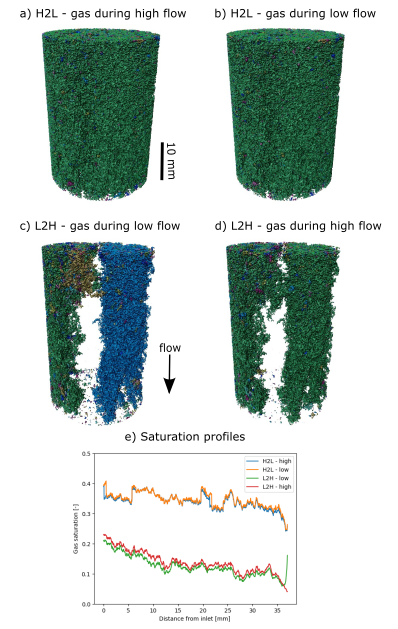}
	\end{center}
	\caption{a-d) the gas distribution across the core during high flow and low flow for both injection scenarios. Here only gas is shown, with the rock grains and brine transparent. Separate gas ganglia are coloured to show the connectivity of the gas phase. e) slice-averaged saturation profiles during high and low flow for both injection scenarios.}
	\label{gas_distribution}
\end{figure}

The gas distribution across the core for the different injection scenarios (H2L and L2H) is shown in Figure \ref{gas_distribution}a-d; only the gas is shown, with the rock grains and brine transparent. Each color corresponds to a different ganglion of gas to highlight gas connectivity across the core. The slice averaged saturation across the core during high flow and low flow for both injection scenarios is quantified in Figure \ref{gas_distribution}e. 

For the H2L experiment, the decrease in flow rate does not change the slice averaged saturation, as shown in Figure \ref{gas_distribution}e. However, this does not mean that no fluid rearrangement occurred, as discussed in the next section. For the L2H experiment, the increase in flow rate only increases the gas saturation slightly. In Figure \ref{gas_distribution} c and d, there is a marked increase in connectivity of the gas phase during high flow, heavily linked to the heterogeneous band in the middle of the sample. This is discussed in the next section.

During low flow in the L2H experiment, a flow path for the gas across the core was established (see Figure \ref{gas_distribution}c). When the flow rate was increased, the saturation change was small, leading to a lower saturation at the end of drainage than the H2L experiment (see Figure \ref{gas_distribution} e), even though the maximum flow rate, and the volume injected at both flow rates was the same for both injection scenarios. Thus, once a flow path is established, a change in pore occupancy is less likely to occur, for a given flow rate. The pressure generated by the high-rate injection can dissipate more easily if a flow path is already established. Therefore, there is a smaller difference between the two fluids present, hence less displacement. 

Overall, the largest pore volume utilisation was during the H2L experiment. As the volume injected was the same, this implies that, if upscaled, that a H2L injection regime will minimise plume spread for a given volume of gas, due to a more compact displacement. Once a pathway across the pore space is established, it is more difficult to increase pore volume utilisation. This implies that inreasing the flow rate with time leads to an irreversible reduction in the pore space accessible for the gas. 

\subsection{The impact of injection scenario on gas connectivity during drainage}
\begin{figure}[!htb]
	\begin{center}
		\centering
		\includegraphics*[angle=0, width=0.8\linewidth]{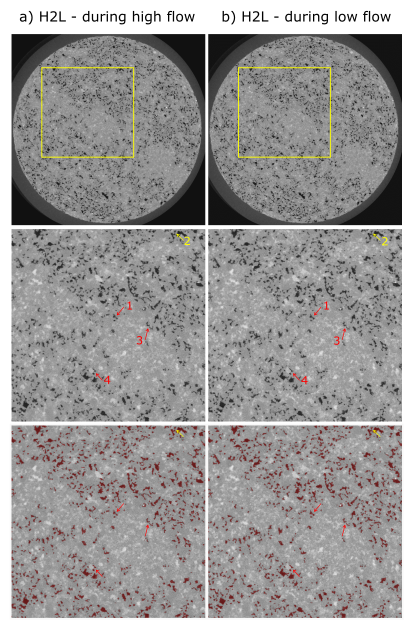}
	\end{center}
	\caption{Fluid distribution across a slice for a) during high flow in the H2L experiment and b) during low flow in the H2L experiment. The yellow square is blown up to highlight some displacement events that occur. Red arrows highlight connection events, and yellow arrows highlight disconnection events. These events are enlarged in Figure \ref{h2l_zoom}. The segmentation is also shown to highlight that these dynamics are captured by the segmentation.}
	\label{h2l_drainage_displacements}
\end{figure}

\begin{figure}[!htb]
	\begin{center}
		\centering
		\includegraphics*[angle=0, width=0.7\linewidth]{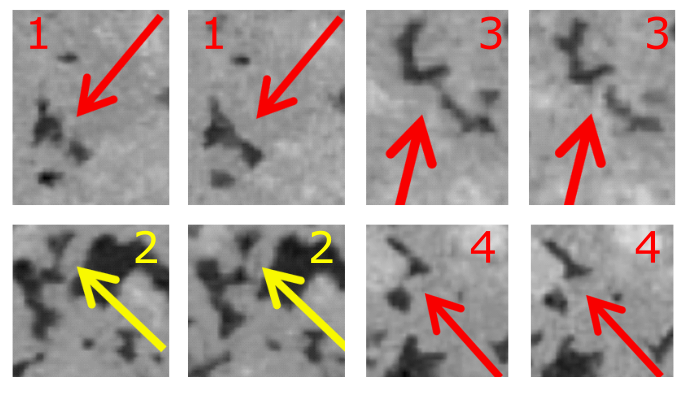}
	\end{center}
	\caption{Enlarged versions of the displacement events highlighted in Figure \ref{h2l_drainage_displacements}. 1, 3 and 4 are snap-off events, and 2 is a connection event.}
	\label{h2l_zoom}
\end{figure}

\begin{figure}[!htb]
	\begin{center}
		\centering
		\includegraphics*[angle=0, width=0.7\linewidth]{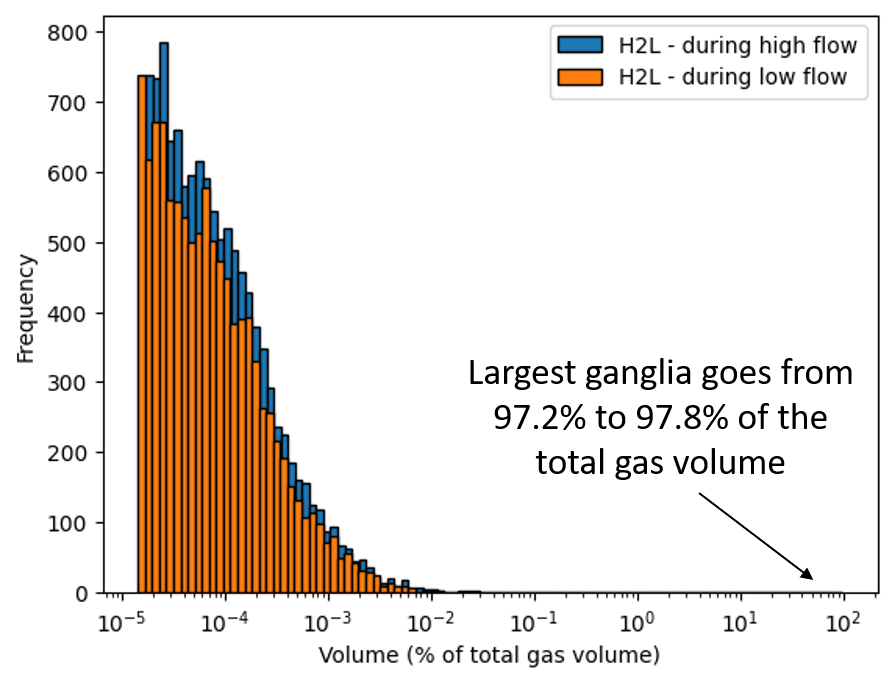}
	\end{center}
	\caption{Ganglia size-frequency distribution plot during high flow and low flow for the H2L experiment.}
	\label{ganglia_size_freq}
\end{figure}

For the H2L experiment, the decrease in flow rate causes a number of small rearrangements of the fluids in the pore space. Both snap-off and connection of the gas is observable, as shown in Figure \ref{h2l_drainage_displacements} and Figure \ref{h2l_zoom}. These events are small, with the largest ganglion contributing 97.2\% of the total gas volume, but collectively influence the connectivity of the gas. As shown in Figure \ref{ganglia_size_freq}, the smaller ganglia decrease in frequency for all ganglion sizes, with the largest ganglion increasing in volume to 97.8\% of the total gas volume as a result. The total number of ganglia reduces by approximately 13\%, from 12,442 to 10,835. While saturation changes across the core are small (all within 2\%, see Figure \ref{gas_distribution}e), the connectivity in the small ganglia changes significantly. Previous research suggested that lowering the flow rate would favour snap-off, which would encourage trapping \cite{hughes2000pore, herring2015efficiently}. While we observe snap-off events, the global trend is for the smaller ganglia to connect to the main flow pathway, reducing the volume of disconnected gas at the end of drainage.

\begin{figure}[!htb]
	\begin{center}
		\centering
		\includegraphics*[angle=0, width=0.7\linewidth]{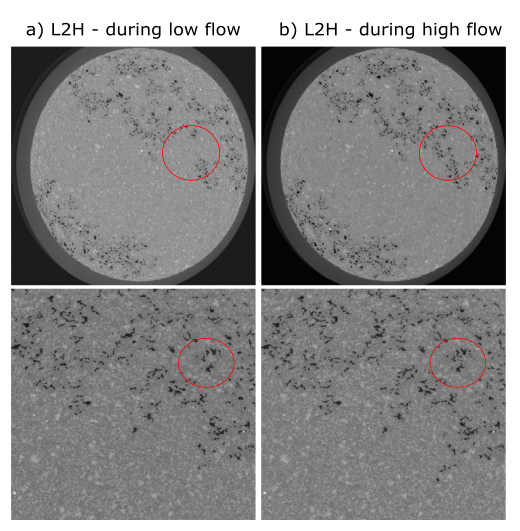}
	\end{center}
	\caption{Fluid distribution across a slice for a) during low flow in the L2H experiment and b) during high flow in the L2H experiment. The red circles highlight regions that become filled during high flow.}
	\label{l2h_drainage_displacements}
\end{figure}

\begin{figure}[!htb]
	\begin{center}
		\centering
		\includegraphics*[angle=0, width=0.7\linewidth]{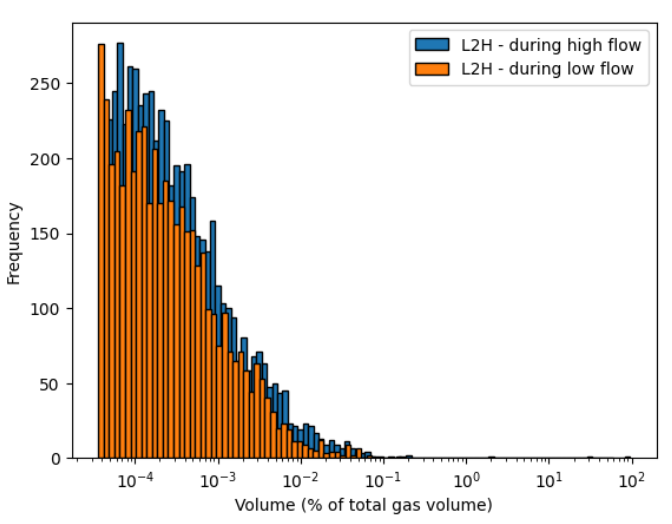}
	\end{center}
	\caption{Ganglia size-frequency distribution plot during high flow and low flow for the L2H experiment.}
	\label{ganglia_size_freq_l2h}
\end{figure}

For the L2H experiment, the increase in flow rate causes some filling events, which greatly increases the connectivity of the gas phase, shown qualitatively in Figure \ref{gas_distribution} c-d and Figure \ref{l2h_drainage_displacements}. The number of ganglia decreases from 5,922 to 4,771, a change of 24\%, relative to a 9\% increase in saturation. The largest ganglion increases in volume, from 58.9\% of the total gas volume, to 94.9\% of the total gas volume. This large increase is due to the connection of the two sides either side of the unoccupied band. 

The increase in flow rate also increases the number of smaller ganglia, as shown in Figure \ref{ganglia_size_freq_l2h}. Thus, there is filling that increases connectivity, and also filling that does not improve connectivity, with the filling events improving connectivity outweighing the filling events that do not. While connectivity increases, there is a band completely unoccupied by gas in the middle of the sample, even at the end of drainage, observable in Figure \ref{gas_distribution} c-d. During the H2L experiment, this band is filled with gas.

\subsection{The impact of injection scenario on trapping efficiency}
\begin{figure}[!htb]
	\begin{center}
		\centering
		\includegraphics*[angle=0, width=0.7\linewidth]{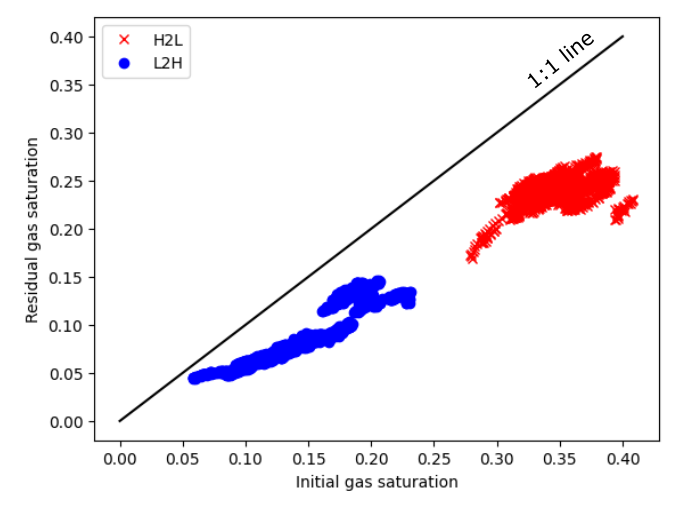}
	\end{center}
	\caption{Trapping curves for the H2L and L2H experiment. The initial gas saturation is the gas saturation at the end of drainage, and the residual gas saturation is the gas saturation at the end of imbibition.}
	\label{trapping}
\end{figure}

\begin{figure}[!htb]
	\begin{center}
		\centering
		\includegraphics*[angle=0, width=0.8\linewidth]{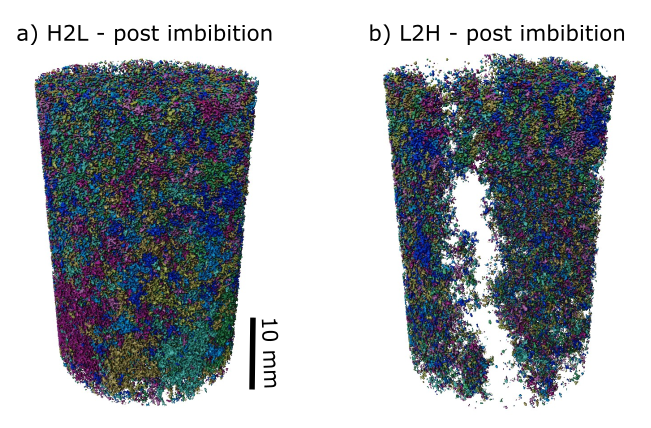}
	\end{center}
	\caption{Gas distribution a) post imbibition for the H2L experiment, and b) post imbibition for the L2H experiment. The brine and rock grains are transparent, and each gas ganglion is coloured differently to highlight the connectivity of the gas phase.}
	\label{post_imbibition_gas_distribution}
\end{figure}

\begin{figure}[!htb]
	\begin{center}
		\centering
		\includegraphics*[angle=0, width=0.9\linewidth]{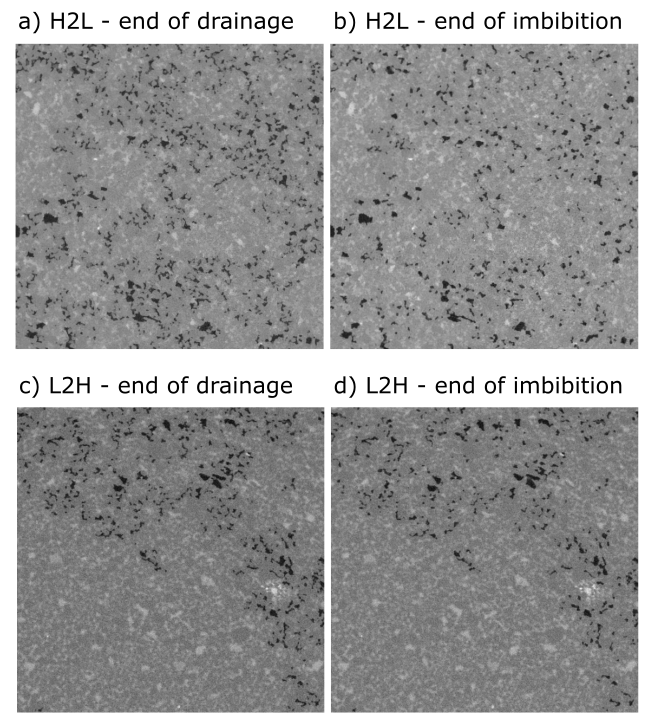}
	\end{center}
	\caption{Fluid distribution across a slice for a) H2L at the end of drainage, b) H2L at the end of imbibition, c) L2H at the end of drainage, and d) L2H at the end of imbibition.}
	\label{post_imbibition_slices}
\end{figure}

\begin{figure}[!htb]
	\begin{center}
		\centering
		\includegraphics*[angle=0, width=0.7\linewidth]{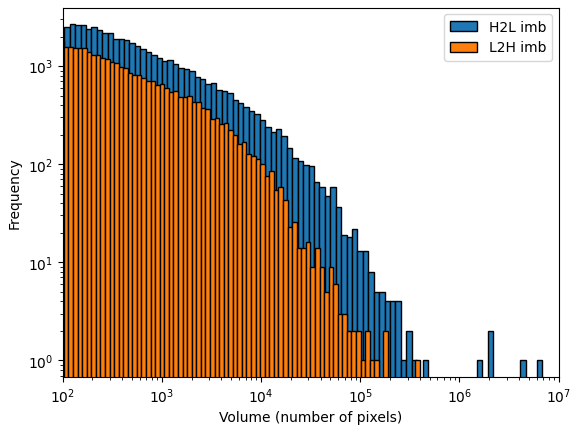}
	\end{center}
	\caption{Ganglia size-frequency distribution plot at the end of imbibition for the H2L and L2H experiments}
	\label{ganglia_size_freq_imb}
\end{figure}

The initial gas saturation (the saturation at the end of drainage) is plotted against the residual gas saturation (the saturation at the end of imbibition) to obtain the trapping efficiency for both experiments. This is shown in Figure \ref{trapping}. For the L2H experiment, the initial gas saturation is lower, but trapping efficiency is similar for both experiments. Thus, the H2L experiment presents a more favourable trapping scenario. 

As can be observed qualitatively in Figure \ref{post_imbibition_gas_distribution} and Figure \ref{post_imbibition_slices}, there are larger gas ganglia post imbibition for the H2L imbibition experiment. This is quantified in Figure \ref{ganglia_size_freq_imb}, with some much larger ganglia present at the end of imbibition for the H2L experiment (over an order of magnitude larger than the largest ganglia present at the end of imbibition for the L2H experiment). The largest ganglia for the H2L experiment contributes 5\% of the total gas volume, compared to 2\% of the total gas volume for the L2H experiment, so the gas ganglia are relatively disconnected in both cases. For the H2L experiment, there are also more smaller disconnected regions.

\section{Conclusions}
In this work, we show that the injection scenario influences the extent of the pore volume accessed by the gas, and the amount of residual trapping. Starting at a high flow rate, which subsequently decreases, causes a larger volume of the pore space to be accessed by the gas. While the trapping efficiency is slightly higher in the experiment where the flow rate is increased, the water channels through part of the sample not accessed by the gas. This means the control on trapping efficiency could be dependent on the heterogeneity of the sample. Further work should explore different lithologies, so that the role of heterogeneity could be assessed. Variations in pore space accessibility could be less if the sample is more homogeneous, or when averaged over a larger volume, which would influence residual trapping. However, this needs to be explored in more detail. 

The high to low injection scenario is the most favourable in terms of total amount of gas trapped related to gas injected. However, there are other implications that control the injection of gas into the subsurface, such as minimising the maximum pressure to prevent fracturing. This must be kept in mind when developing an injection protocol for subsurface storage. While starting at a lower flow rate is associated with a lower risk for fracturing, increasing the flow rate at a later time does not significantly change the relative amount of the pore volume accessed, or the average saturation. This would translate at the field scale to the gas plume spreading out further and slower stabilisation times.

\section{Acknowledgements}
Catherine Spurin and Hamdi Tchelepi acknowledge support from the GeoCquest consortium. 
Sharon Ellman is a PhD Fellow with the Research Foundation – Flanders (FWO) and acknowledges its support under grant 1182822N. T. Bultreys holds a senior postdoctoral fellowship from the Research Foundation-Flanders (FWO) under Grant No. 12X0922N. This research also received funding from the Research Foundation–Flanders under grant G051418N, G004820N and the UGent BOF funding for the Centre of Expertise UGCT (BOF.EXP.2017.0007).

\bibliography{refs}

\end{document}